\documentclass[gmd]{copernicus}

\newcolumntype{L}{>{\displaystyle}l} 
\newcolumntype{C}{>{\displaystyle}c} 
\newcolumntype{R}{>{\displaystyle}r} 
\def\bn{\mathbf{n}}
\def\bt{\mathbf{t}}
\def\div{\nabla\cdot}

\def\bbsigma{\mathbb{\sigma}}
\def\bv{\mathbf{v}}

\def\fatg{\mathbf g}

\def\calH{\mathcal{H}}
\def\calL{\mathcal{L}}
\def\calN{\mathcal{N}}
\def\tr{\text{tr}}

\def\fatx{\mathbf x}

\def\bbtau{\mathbf{\mathbb{\tau}}}

\def\calO{\mathcal{O}}
\def\calT{\mathcal{T}}
\def\calE{\mathcal{E}}
\def\bu{\mathbf{u}}
\def\dx{\ \text{d}\fatx}
\def\ds{\ \text{d}s}

\begin{document}

\nolinenumbers
\title{A full Stokes subgrid model for simulation of grounding line migration in ice sheets}

\Author[1]{Gong}{Cheng}
\Author[1]{Per}{L\"{o}tstedt}
\Author[1]{Lina}{von Sydow}

\affil[1]{Department of Information Technology, Uppsala University, P. O. Box 337, SE-75105 Uppsala, Sweden}

\runningtitle{{Full Stokes subgrid model for GL}}
\runningauthor{G. Cheng et al.}
\correspondence{Gong Cheng (cheng.gong@it.uu.se)}

\received{}
\pubdiscuss{} 
\revised{}
\accepted{}
\published{}


\firstpage{1}
\maketitle

\begin{abstract}
The full Stokes equations are solved by a finite element method for simulation of large ice sheets and glaciers. The simulation is particularly sensitive to the discretization of
the grounding line which separates the ice resting on the bedrock and the ice floating on water and is moving in time. The boundary conditions at the ice base are enforced by Nitsche's method and
a subgrid treatment of the elements in the discretization close to the grounding line. Simulations with the method in two dimensions for an advancing and a retreating grounding line
illustrate the performance of the method. It is implemented in the two dimensional version of the open source code Elmer/ICE.  
\end{abstract}


\introduction
Simulation with ice sheet models is a tool to assess the future sea-level rise (SLR) due to melting of continental ice sheets and glaciers \cite{Hanna13} and to reconstruct the
ice sheets of the past \cite{DeConto16, Stokesetal} for comparison with measurements and validation of the models.
In the models, the predictions are particularly sensitive to the numerical treatment of the grounding line (GL) \cite{Durand15}.
The GL is the line where the ice sheet leaves the solid bedrock and becomes an ice shelf floating on water driven by buoyancy.
It is important to know the GL position to be able to quantify the ice discharge into the sea and as an indicator if the ice sheet is advancing or retreating \cite{Konrad18}.
The distance that the GL moves may be long over palaeo time scales.
It is shown in \cite{Kingslake18} that the GL has retreated several hundred km on West Antarctica during the last 11,500 years and then advanced again 
after the isostatic rebound of the bed. The sensitivity, long time intervals, and long distances require a careful treatment of the GL neighborhood 
by the numerical method to discretize the model equations.

The most accurate ice model is based on the full Stokes (FS) equations. A simplification of the FS equations by integrating in the depth of the ice 
is the shallow shelf (or shelfy stream) approximation (SSA) \cite{SSA}. The computational advantage with SSA is that the dimension of the problem is reduced by one. 
It is often used for simulation of the interaction between a grounded ice sheet and a marine ice shelf. Several other simplifications exist with the
same advantages as the SSA but with slightly different solutions.
Another simplification is the shallow ice approximation (SIA) suitable for ice sheets where vertical shear stresses determine the ice flow \cite{WGH99}.

When the ice rests on the ground and is affected by frictional forces on the bed, the ice flow is dominated by vertical shear stresses. 
Longitudinal stresses are dominant when the ice is floating on water. The GL is in the transition zone with a gradual change of the stress field.
A SSA model for a two dimensional (2D) ice is analyzed in \cite{Schoof07} where there is a switch in the friction coefficient at the GL from being positive in the grounded ice to zero in the
floating ice. The stability of steady state GL solutions depends on the geometry of the slope, see \cite{Schoof07}. It is stable in a downward slope and unstable in an upward slope.
In the zone between the grounded ice and the floating ice, it is necessary to use the FS equations \cite{Docquier11, Schoof11, SchoofHindmarsh, Wilchinsky00} unless the ice is
moving rapidly on the ground with low basal friction and the SSA equations are accurate both upstream and downstream of the GL. The solution to the linearized FS equations close to the GL is
investigated using perturbation theory in \cite{Schoof11}. The effect of perturbations in the topography and the friction coefficient on the surface velocity and height 
is studied in \cite{CGPL19a}.
The sensitivity to the perturbations increases close to the GL because the velocity of the ice increases and the thickness decreases there.

The evolution of the GL in simulations is sensitive to the ice model, the basal friction model, and numerical parameters.
In a major effort MISMIP \cite{MISMIP3d, MISMIP}, different ice models and implementations solve the same ice flow problems and the predicted GL steady state and
transient GL motion are compared. The results depend on the model equations and the mesh resolution \cite{MISMIP3d}.
The prediction of the GL and the SLR is different for different ice equations such as FS and SSA also in \cite{Pattyn13}. 
Including equations with vertical shear stress at the GL such as the FS equations seems to be crucial.
The friction laws at the ice base depend on the effective pressure, the basal velocity, and distance to the GL in different combinations 
in \cite{Brondex17, Gagliardini15, Gladstone17, Leguy14}. 
The GL position and the SLR vary considerably depending on the choice of friction model. Given the friction model, the results are sensitive to its model parameters too \cite{Gong17}.

Parameters in the numerical methods also influence the GL migration. 
It is observed in \cite{Durand09a} that the mesh resolution along the ice bed has to be fine to obtain reliable solutions with FS in GL simulations. The GL is then
located in a node of the fixed mesh. 
A mesh size below 1~km is necessary in \cite{Larour19} to resolve the features at the GL. 
The SIA and SSA equations model the ice close to the GL in \cite{Docquier11}. 
The transient response of the GL is compared with the FS equations and adaptive meshes in 2D and the SSA equations in \cite{Drouet13}. 
The flotation condition determines where the GL is in \cite{Docquier11, Drouet13}. It is based on Archimedes' principle for an ice column immersed in water.
Another adaptive mesh method is developed for the SSA equations in 1D in \cite{Gladstone10a}. The accuracy of the method is evaluated in simulations of the GL migration. Adaptive meshes for a finite volume discretization of an approximation of the FS equations are employed in \cite{Cornford13} to study the 
GL retreat and loss of ice in West Antarctica. The FS solutions of benchmark problems
in \cite{MISMIP3d} computed by FEM implementations in Elmer/ICE \cite{ElmerDescrip} and FELIX-S \cite{LJGPR} are compared in \cite{ZPJLBDG}. 
The differences between the codes are attributed to different treatment of a friction parameter at the GL and different assignment of grounded and floating nodes and element faces.

A subgrid model introduces an inner structure in the discretization element or mesh volume where the GL is.
Such a model for the GL is tested in \cite{Gladstone10b} for the 1D SSA equation where the flotation condition for the ice defines the position of the GL.
The GL migration is determined by the 2D SSA equations discretized by the finite element method (FEM) in \cite{Seroussi14}. Subgrid models at the GL are compared to
a model without an internal structure in the element. The conclusion is that sub-element parameterization is necessary. 
A shallow approximation to FS with subgrid modeling on coarse meshes is compared to FS in \cite{Feldmann14} with similar results for the GL migration.
Subgrid modeling and adaptivity are compared in \cite{Cornford16} for a vertically integrated model. The stability of the GL in solutions with FS and fine meshes in 2D are 
compared in \cite{Durand09b} to the theory in \cite{Schoof07} with good agreement. A fine mesh resolution is necessary for converged 
GL positions with FS in \cite{Durand09b, Durand09a}. The purpose of a subgrid model is to avoid such fine meshes.   

The fine mesh resolution needed in GL simulations with the FS equations would require large computational efforts in 3D to solve the equations in long time intervals.
Since the GL moves long distances in palaeo simulations, a dynamic mesh refinement and coarsening of the mesh following the GL is necessary.
The alternative pursued here is to introduce a subgrid modeling with FEM in the mesh elements where the GL is located and keep the mesh size coarser. 
The subgrid model is restricted to one element in a 2D ice and is therefore computationally inexpensive. In an extension to 3D, the subgrid model would be applied along a 1D line of elements in 3D. 
The results with numerical modeling will always depend on the mesh resolution but can be more or less sensitive to the mesh spacing and time steps.
Our subgrid modeling is aiming at improving the accuracy in GL simulations for a static mesh size.

We solve the FS equations in 2D with the Galerkin method implemented in Elmer/ICE \cite{ElmerDescrip}. The boundary conditions are imposed by Nitsche's method in the weak formulation of the equations \cite{Nitsche, Reusken17, Urquiza14}. 
The linear Stokes equations are solved in \cite{chouly2017overview} with Nitsche's treatment of the boundary conditions. 
They solve the equations for the displacement but here we solve for the velocity using similar numerical techniques to weakly impose the Dirichlet boundary conditions.
A subgrid discretization is proposed and tested for the element where the GL is located. The position of the GL within the element is determined by theory developed for the linearized FS in \cite{Schoof11}. 

The paper is organized as follows. Section \ref{sec:mathmodel} is devoted to the presentation of the mathematical model of the ice sheet dynamics. 
In  Sect. \ref{sec:discr} the numerical discretization is presented while the subgrid modeling around the GL is found in Sect. \ref{sec:subgrid}. 
We present the numerical results in Sect. \ref{sec:results}. The extension to 3D is discussed in Sect. \ref{sec:disc} and finally some conclusions are drawn in Sect. \ref{sec:concl}.


\section{Ice model} \label{sec:mathmodel}

\subsection{The full Stokes (FS) equations}
\label{sec:Stokes}
We use the FS equations in 2D with coordinates $\fatx=(x, z)^T$ for modeling of the flow of an ice sheet \cite{Hutter83}.
These nonlinear partial differential equations (PDEs) in the interior of the ice $\Omega$ are given by 
\begin{equation}
\begin{cases}
    \div\bu=0,\\
    - \div{\mathbf{\bbsigma}} =\rho \mathbf{g},
 \end{cases}
  \label{eq:FS}
\end{equation}
  where the stress tensor is ${\mathbf{\bbsigma}} = 2\eta(\bu)\bbtau(\bu)-p\mathbb{I}$. The symmetric strain rate tensor is defined by
\begin{equation}\label{eq:taudef}
  \bbtau(\bu)=\frac{1}{2}(\nabla\bu+\nabla\bu^T)=\left(\begin{array}{cc}\tau_{11}&\tau_{12}\\\tau_{12}&\tau_{22}\end{array}\right),
\end{equation}
$\mathbb{I}$ is the identity matrix, and the viscosity is defined by Glen's flow law
\begin{equation}\label{eq:visc}
  \eta(\bu)=\frac{1}{2}\left(\mathcal{A}(T^\prime)\right)^{-\frac{1}{n}}\bbtau_e^{\frac{1-n}{n}},\qquad \bbtau_e = \sqrt{\frac{1}{2}\tr(\bbtau(\bu)\bbtau(\bu))}.
\end{equation}

Here ${\bu}=(u, w)^T$ is the vector of velocities, $\rho$ is the density of the ice, $p$ denotes the pressure, and the gravitational acceleration in the $z$-direction is denoted by ${\bf g}$. The rate factor $\mathcal{A}(T^\prime)$ describes how the viscosity depends on the pressure melting point corrected temperature $T^\prime$. For isothermal flow assumed here, the rate factor $\mathcal{A}$ is constant. Finally, $n$ is usually taken to be 3.

\subsection{Boundary conditions}
\begin{figure}[htbp]
\center
  \includegraphics[width=0.45\textwidth]{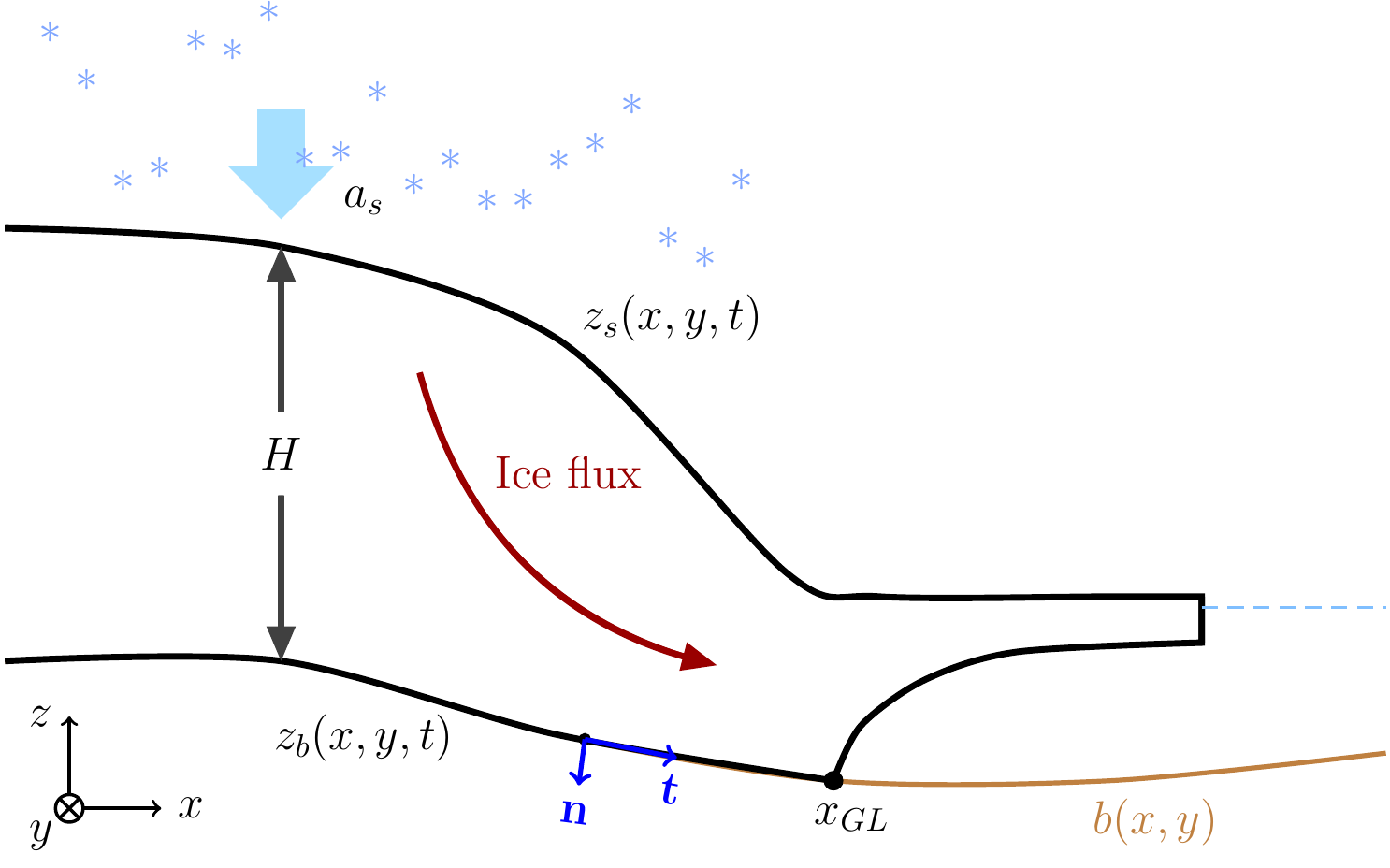} 
  \caption{A two dimensional schematic view of a marine ice sheet.}
\label{fig:ice}
\end{figure}

At the boundary $\Gamma$ of the ice we define the normal outgoing vector $\bn$ and tangential vector $\bt$, see Figure~\ref{fig:ice}.
In a 2D case considered here, $y$ is constant in the figure. The upper boundary is denoted by $\Gamma_s$ and the lower boundary is $\Gamma_b$.
At $\Gamma_s$ and $\Gamma_{bf}$, the floating part of $\Gamma_b$, we have that 
\begin{equation}
{\mathbf{\bbsigma}}{\bn}=\mathbf{f}_s.
\label{eq:bc_s}
\end{equation}
The ice is stress-free at $\Gamma_s$, $\mathbf{f}_s=0$, and $\mathbf{f}_s=-p_w\bn$ at the ice/ocean interface $\Gamma_{bf}$ where $p_w$ is the
water pressure. Let
\[
   \mathbf{\bbsigma}_{\bn\bt}=\bt\cdot\mathbf{\bbsigma}\bn,\; \mathbf{\bbsigma}_{\bn\bn}=\bn\cdot\mathbf{\bbsigma}\bn,\; u_\bt=\bt\cdot\bu.
\]
Then for the slip boundary $\Gamma_{bg}$, the part of $\Gamma_b$ where the ice is grounded, we have a friction law for the sliding ice
  \begin{equation}
      {\mathbf{\bbsigma}}_{\bn\bt} + \beta(\bu,\fatx) u_\bt=0,\quad u_\bn=\bn\cdot\bu=0, \quad -{\mathbf{\bbsigma}}_{\bn\bn}\geq p_w.       \label{eq:BCGI}
    \end{equation} 
The type of friction law is determined by the friction coefficient $\beta$.
There is a balance between ${\mathbf{\bbsigma}}_{\bn\bn}$ and $p_w$ at $\Gamma_{bf}$ and the contact is friction-free, $\beta=0$,
            \begin{equation}
           {\mathbf{\bbsigma}}_{\bn\bt} = 0, \qquad -{\mathbf{\bbsigma}}_{\bn\bn}= p_w.
              \label{eq:BCFI}
          \end{equation}
The GL is located where the boundary condition switches from $\beta>0$ and $u_\bn=0$ on $\Gamma_{bg}$ to $\beta=0$ and a free $u_\bn$ on $\Gamma_{bf}$. In 2D,
the GL is the point $(x_{GL}, z_{GL})$ between $\Gamma_{bg}$ and $\Gamma_{bf}$. 

With the ocean surface at $z=0$, $p_w=-\rho_w g z_b$ where $\rho_w$ is the density of sea water, $z_b$ is the $z$-coordinate of $\Gamma_b$, and $g$ is the gravitation constant.

\subsection{The free surface equations}
\label{sec:height}

The boundaries $\Gamma_s$ and $\Gamma_b$ are time-dependent and move according to two free surface equations. The boundary $\Gamma_{bg}$ follows the
fixed bedrock with coordinates $(x, b(x))$.

The $z$-coordinate of the free surface position $z_s(x,t)$ at $\Gamma_s$ (see Fig. \ref{fig:ice}) is the solution of an advection equation
\begin{equation}
  \frac{\partial z_s}{\partial t}+u_s \frac{\partial z_s}{\partial x}-w_s=a_s,
\label{eq:freeSurface}
\end{equation}
where $a_s$ denotes the net surface accumulation/ablation of ice and ${\bu}_s=(u_s, w_s)^T$ the velocity at the free surface in contact with the atmosphere. Similarly, the $z$-coordinate for the lower surface $z_b$ of the floating ice at $\Gamma_{bf}$ satisfies
\begin{equation}
  \frac{\partial z_b}{\partial t}+u_b \frac{\partial z_b}{\partial x}-w_b=a_b,
\label{eq:lowerSurface}
\end{equation}
where $a_b$ is the net accumulation/ablation at the lower surface and ${\bu}_b=(u_b, w_b)^T$ the velocity of the ice at $\Gamma_{bf}$. On $\Gamma_{bg}$, $z_b=b(x)$.

The thickness of the ice is denoted by $H=z_s-z_b$ and depends on $(x, t)$. 

\subsection{The solution close to the grounding line}
\label{sec:GLsol}

The 2D solution of the FS equations in Eq. \eqref{eq:FS} with a constant viscosity, $n=1$ in Eq. \eqref{eq:visc}, is expanded in small parameters in \cite{Schoof11}. The solutions in different
regions around the GL are connected by matched asymptotics. Upstream of the GL at the bedrock, $x<x_{GL}$, the leading terms in the expansion satisfy
a simple equation in scaled variables close to the GL. Across the GL, $u$, the flux of ice $uH$, and the depth integrated normal or longitudinal stress $\tau_{11}$ in Eq. \eqref{eq:taudef} are continuous.
By adding higher order terms, it is shown that the upper surface slope is continuous and Archimedes' flotation condition
\begin{equation}
  H\rho=-z_b\rho_w
\label{eq:flot}
\end{equation}
is not satisfied immediately downstream of the GL.
A rapid variation in the vertical velocity $w$ in a short interval at the GL causes oscillations in the upper surface as observed also in FS simulations in \cite{Durand09b}.

In \cite[Ch. 4.3]{Schoof11}, the solution to the FS in 2D is expanded in two parameters $\nu$ and $\epsilon$. The aspect ratio of the ice $\nu$ is the quotient
between a typical scale of the height of the ice $\calH$ and a length scale $\calL$, $\nu=\calH/\calL$, and $\epsilon$
is $\nu$ times the quotient between the longitudinal and the shear stresses $\tau_{11}$ and $\tau_{12}$ in Eq. \eqref{eq:taudef}. 
If $\nu^{5/2}\ll \epsilon\ll 1$ then in a boundary layer close to the GL and $x<x_{GL}$ the 
leading terms in the solution in scaled variables satisfy
\begin{equation}
  \tau_{22}-p={\sigma}_{22}=\rho g(z-z_s).
\label{eq:perturbsol}
\end{equation}
On floating ice $\tau_{22}-p+p_w=0$ and the flotation criterion Eq. \eqref{eq:flot} is fulfilled, and on the bedrock $\tau_{22}-p+p_w<0$, see Eq. \eqref{eq:bc_s} and \eqref{eq:BCFI}.

Introduce the notation
\begin{equation}
  \chi(x, z)=\tau_{22}-p+p_w=\rho g(z-z_s(x))-\rho_wgz_b(x),
\label{eq:chidef}
\end{equation}
and approximate $z_s$ and $z_b$ linearly in $x$ in the vicinity of $x_{GL}$ and let $H_{bw}$ be the thickness of the ice below the water surface. 
Then 
\begin{equation}
  \chi(x,z_b)=-g(\rho H-\rho_w H_{bw})
\label{eq:chidef2}
\end{equation}
is linear in $x$. If $x<x_{GL}$ then $\chi<0$ in the neighborhood of $x_{GL}$ on $\Gamma_{bg}$ and if $x>x_{GL}$ then $\chi=0$ and Eq. \eqref{eq:flot} holds true
on $\Gamma_{bf}$.
In numerical experiments with the linear FS $(n=1)$ in \cite{NoWi08}, $\chi(x, z_b)$ in the original variables varies linearly in $x$ for $x<x_{GL}$. 
In Sect. \ref{sec:subgrid}, $\chi(x, z_b)$ is used to estimate the GL position.

\section{Discretization by FEM}
\label{sec:discr}

In this section we state the weak form of Eq. \eqref{eq:FS} and introduce the spatial FEM discretization used for Eq. \eqref{eq:FS} and give the time-discretization of Eq. (\ref{eq:freeSurface}) and (\ref{eq:lowerSurface}). 

\subsection{The weak form of the FS equations}
\label{sec:weakFS}

We start by defining the mixed weak form of the FS equations.
Introduce $k=1+1/n$, $k^*=1+n$ and the spaces
    \begin{equation}\label{eq:spaces}
        {\boldsymbol{V}}_k= \{\bv:\bv\in (W^{1,k}(\Omega))^2 \}, \quad
        Q_{k^*}= \{q: q\in L^{k^*}(\Omega)\},
    \end{equation}
see, e.g. \cite{chen2013well,martin2014four}.
The weak solution $(\bu,p)$ of Eq. \eqref{eq:FS} is obtained as follows.
Find $(\bu,p)\in{\boldsymbol{V}}_k\times Q_{k^*}$ such that for all $(\bv,q)\in \boldsymbol{V}_k\times Q_{k^*}$ the equation
  \begin{equation}\label{eq:FSweakform}
    A((\bu, p), (\bv,q)) +B_\Gamma(\bu, p, \bv) + B_\calN(\bu, \bv, q) =F(\bv),
  \end{equation} 
 is satisfied, where 
  \begin{equation}
    \begin{split}
      &A((\bu, p), (\bv,q))=\int_{\Omega}2\eta(\bu)\bbtau(\bu) :\bbtau(\bv)\ \dx  - b(\bu,q) -b(\bv, p),\\
      &b(\bu,q) = \int_{\Omega}q\div \bu\ \dx,\\
      &B_\Gamma(\bu, p, \bv)= \int_{\Gamma_{bg}} \left(-\mathbf{\bbsigma}_{\bn\bn}(\bu, p) \bn\cdot\bv +\beta  \bu \cdot\bv\right)\,\ds,\\
      &   B_\calN(\bu, \bv, q) = -\int_{\Gamma_{bg}} \mathbf{\bbsigma}_{\bn\bn}(\bv, q) \bn\cdot\bu \ds +\gamma_0\int_{\Gamma_{bg}}\frac{1}{h}(\bn\cdot\bu)(\bn\cdot\bv) \ds,\\
      &F(\bv)=\int_{\Omega}\rho\fatg\cdot\bv\ \dx -\int_{\Gamma_{bf}} p_w\bn\cdot\bv\,\ds. \nonumber  
    \end{split}
  \end{equation}

The last term in $B_\calN$ is added in the weak form in Nitsche's method \cite{Nitsche} to impose the Dirichlet condition $u_\bn=0$ weakly on $\Gamma_{bg}$. 
It can be considered as a penalty term. 
The size of the positive parameter $\gamma_0$  depends on the application and $h$ is a measure of the mesh size on $\Gamma_b$. 
The first term in $B_{\calN}$ symmetrizes the boundary term $B_\Gamma+B_\calN$ on $\Gamma_{bg}$ and vanishes when $u_\bn=0$.


\subsection{The discretized FS equations}
\label{sec:discrFS}

We employ linear Lagrange elements with Galerkin Least Square (GLS) stabilization \cite{franca1992stabilized,helanow2018stabilized} to avoid spurious oscillations in the pressure using the standard setting in Elmer/ICE \cite{ElmerDescrip} approximating solutions in the spaces ${\boldsymbol{V}}_k$ and $ M_{k^*}$ in Eq. \eqref{eq:spaces}.

The mesh is constructed from a footprint mesh on the bottom surface and then extruded with the same number of layers in the vertical direction according to the thickness of the ice.
To simplify the implementation in 2D, the footprint mesh on the bottom surface consists of $N+1$ nodes $x_i,\; i=0,\ldots,N,$ with a constant mesh size $\Delta x$.

In general, the GL is somewhere in the interior of an element $\calE_i=[x_i,\, x_{i+1}]$ and it crosses the element boundaries as it moves forward in the advance phase and backward in the retreat phase of the ice. The advantage with Nitsche's way of formulating the boundary conditions is that if $x_{GL}\in \calE_i$ then the
boundary integral over $\calE_i$ can be split into two parts in Eq. \eqref{eq:FSweakform} such that $[x_i,\, x_{GL}]\in \Gamma_{bg}$ and $[x_{GL},\, x_{i+1}]\in \Gamma_{bf}$ as follows
 \begin{equation}\label{eq:Nitscheint}
\begin{split}
   &{\int_{\calE_i} B_\Gamma +B_\calN\, \ds} \\
   =& {\int_{[x_i,\, x_{GL}]}-({\mathbf{\bbsigma}}_{\bn\bn}(\bu, p) \bn\cdot\bv+{\mathbf{\bbsigma}}_{\bn\bn}(\bv, q) \bn\cdot\bu)} \\
   &{+ \beta\bu\cdot\bv+\frac{\gamma_0}{h}(\bn\cdot\bu)(\bn\cdot\bv)\, \ds+\int_{[x_{GL},\, x_{i+1}]} p_w\bn\cdot\bv\, \ds.}
\end{split}
  \end{equation}
There is a change of boundary conditions in the middle of the element $\calE_i$ where the GL is located. With a strong formulation of $u_\bn=0$ the basis functions in ${\boldsymbol{V}}_s$ share
this property and the condition changes from the grounded node $x_i$ where the basis function satisfies $u_\bn=0$ and the floating node at $x_{i+1}$ with a free $u_\bn$ without 
taking the position of the GL inside $\calE_i$ into account.

The resulting system of non-linear equations form a nonlinear complementarity problem \cite{CKPS}. The distance $d$ between the base of the ice and the bedrock at time $t$ and at
$x$ is $d=z_b(x, t)-b(x)\ge 0$. If $d>0$ on $\Gamma_{bf}$ then the ice is not in contact with the bedrock and $\mathbf{\bbsigma}_{\bn\bn}+p_w=0$ and if $\mathbf{\bbsigma}_{\bn\bn}+p_w<0$ on $\Gamma_{bg}$ then the ice and the bedrock are in contact and $d=0$. Hence, the
complementarity relation in the vertical direction is
 \begin{equation}\label{eq:complv}
\begin{array}{ll}
   z_b(x, t)-b(x)\ge 0,\; \mathbf{\bbsigma}_{\bn\bn}+p_w\le 0,\\ 
  (z_b(x, t)-b(x))(\mathbf{\bbsigma}_{\bn\bn}+p_w)=0\;\textrm{on}\;\Gamma_b.
\end{array}
  \end{equation}
The contact friction law is such that $\beta>0$ when $x<x_{GL}$ and  $\beta=0$ when $x>x_{GL}$. The complementarity relation along the slope at $x$ is then the non-negativity of $d$ and 
 \begin{equation}\label{eq:compls}
   \beta\ge 0,\; \beta(x, t)(z_b(x, t)-b(x))=0\;\textrm{on}\;\Gamma_b.
  \end{equation}
In particular, these relations are valid at the nodes $x=x_j$, $j=0,1,\dots,N$.

The complementarity condition also holds for $u_\bn$ and $\sigma_{\bn\bn}$ such that
 \begin{equation}\label{eq:complu}
\begin{array}{ll}
   \mathbf{\bbsigma}_{\bn\bn}+p_w\le 0,\\ 
  u_\bn(\mathbf{\bbsigma}_{\bn\bn}+p_w)=0\;\textrm{on}\;\Gamma_b,
\end{array}
  \end{equation}
without any sign constraint on $u_\bn$ except for the retreat phase when the ice leaves the ground and $u_\bn<0$.

Similar implementations for contact problems using Nitsche's method are found in \cite{chouly2017overview,chouly2017nitsche}, where the unknowns in the PDEs are the displacement fields
instead of the velocity in Eq. \eqref{eq:FS}.
Analysis in \cite{chouly2017overview} suggests that Nitsche's method for the contact problem can provide a stable numerical solution with an optimal convergence rate.

The nonlinear equations for the nodal values of $\bu$ and $p$ are solved by Newton iterations. The system of linear equations in every Newton iteration is solved iteratively by using the
Generalised Conjugate Residual (GCR) method in Elmer/ICE. The condition on $d_j$ in a node $x_j$ is used for a so called grounded mask, which is computed at each timestep and not changed during the nonlinear iterations.

\subsection{Discretization  of  the advection equations}\label{sec:updlower}

The advection equations for the moving ice boundary in Eq. \eqref{eq:freeSurface} and \eqref{eq:lowerSurface} are discretized in time by a finite difference method and in
space by FEM with linear Lagrange elements for $z_s$ and $z_b$. A stabilization term is added,  making the spatial discretization behave 
like an upwind scheme in the direction of the velocity as implemented in Elmer/ICE.

The advection equations Eq. \eqref{eq:freeSurface} and Eq. \eqref{eq:lowerSurface} are integrated in time by a semi-implicit method of first order accuracy. 
Let $c=s$ or $b$. Then the solution is advanced
from time $t^n$ to $t^{n+1}=t^n+\Delta t$ with the timestep $\Delta t$ by 
\begin{equation}\label{eq:zint}
   z_c^{n+1}=z_c^n+\Delta t(a_c^n-u_c^n \frac{\partial{z_{c}^{n+1}}}{\partial x}+w_c^n).
\end{equation}
The spatial derivative of $z_c$ is approximated by FEM. A system of linear equations is solved at $t^{n+1}$ for $z_c^{n+1}$. This time discretization and its properties are 
discussed in \cite{cheng2017accurate}.

A stability problem in $z_b$ is encountered in the boundary condition at $\Gamma_{bf}$ in \cite{Durand09b}. 
It is solved by expressing $z_b$ in $p_w$ at $\Gamma_{bf}$ with a damping term in \cite{Durand09b}.
An alternative interpretation of the idea in \cite{Durand09b} and an explanation follow below.

The relation between $u_\bn$ and $u_\bt$ at $\Gamma_{bf}$ and $\bu_b=\bu(x, z_b(x))$ is
\begin{equation}
  \bu_b=\left(\begin{array}{c} u_b \\ w_b\end{array}\right)=\left(\begin{array}{c} z_{bx} \\ -1\end{array}\right)\frac{u_\bn}{\sqrt{1+z_{bx}^2}}
        +\left(\begin{array}{c} 1 \\ z_{bx}\end{array}\right)\frac{u_\bt}{\sqrt{1+z_{bx}^2}},
\label{eq:udef}
\end{equation}
where $z_{bx}$ denotes $\partial z_b/\partial x$. Insert $u_b$ and $w_b$ from Eq. \eqref{eq:udef} into Eq. \eqref{eq:lowerSurface} to obtain
\begin{equation}
  \frac{\partial z_b}{\partial t}=a_b-u_\bn\sqrt{1+z_{bx}^2},
\label{eq:zbeq2}
\end{equation}
Instead of discretizing Eq. \eqref{eq:zbeq2} explicitly at $t^n$ with $u_\bn^{n-1}$ to determine $p_w^n$, the base coordinate is updated implicitly
\begin{equation}
  z_{b}^n=z_{b}^{n-1}+\Delta t\left(a_b^n-u_\bn^n\sqrt{1+z_{bx}^2}\right)
\label{eq:zbimpl}
\end{equation}
in the solution of Eq. \eqref{eq:FSweakform}.

Assume that $z_{bx}$ is small.
The timestep restriction in Eq. \eqref{eq:zbimpl} is estimated by considering a 2D slab of the floating ice of width $\Delta x$ and thickness $H$. Newton's law of motion yields
\[
    M \dot{u}_\bn= M g-\Delta x p_w,
\]
where $M=\Delta x(z_s-z_b)\rho$ is the mass of the slab. Divide by $M$, integrate in time for $u_\bn(t^m)$, let $m=n$ or $n-1$, and approximate the integral by the trapezoidal rule for the quadrature to obtain
\[
\begin{split}
      u_\bn(t^m)&=\displaystyle{\int_0^{t^m} g+\frac{g\rho_w}{\rho}\frac{z_b}{z_s-z_b}\,\textrm{d}s} \\
      &\approx \displaystyle{gt^m+\frac{g\rho_w}{\rho}\sum_{i=0}^m\alpha_i\frac{z_b^i}{z_s^i-z_b^i}\Delta t,}\\
\end{split}
\]  
\[
      \alpha_i=0.5, i=0, m,\quad  \alpha_i=1, i=1,\ldots,m-1. 
\]
Then insert $u_\bn^m$ into Eq. \eqref{eq:zbimpl}. All terms in $u_\bn^m$ from timesteps $i<m$ are collected in the sum $\Delta t F^{m-1}$. 
Then Eq. \eqref{eq:zbimpl} can be written
\begin{equation}
  z_{b}^n=z_{b}^{n-1}-\Delta t^2\frac{g\rho_w}{2\rho}\frac{z_b^m}{z_s^m-z_b^m}+\Delta t\left(a_b^n-gt^{m}-\Delta t F^{m-1}\right).
\label{eq:zbimpl2}
\end{equation}
For small changes in $z_b$ in Eq. \eqref{eq:zbimpl2}, the explicit method with $m=n-1$ is stable when $\Delta t$ is so small that
\begin{equation}
 |1-\Delta t^2\frac{g\rho_w}{2H\rho}|\le 1.
\label{eq:stabconde}
\end{equation}
When $H=100$ m on the ice shelf, $\Delta t< 6.1$ s which is far smaller than the stable steps for Eq. \eqref{eq:zint}.
Choosing the implicit scheme with $m=n$, the bound on $\Delta t$ is 
\begin{equation}
1/|1+\Delta t^2\frac{g\rho_w}{2H\rho}|\le 1,
\label{eq:stabcondi}
\end{equation}
i.e. there is no bound on positive $\Delta t$ for stability but accuracy will restrict $\Delta t$.

Much longer stable timesteps are possible at the surface and the base of the ice with a semi-implicit method Eq. \eqref{eq:zint}
and a fully implicit method Eq. \eqref{eq:zbimpl} compared to an explicit method. 
For example, the timestep for the problem in Eq. \eqref{eq:zint} with 1~km mesh size can be up to a couple of months.
Therefore, we use the scheme in Eq. \eqref{eq:zint} for Eq. \eqref{eq:freeSurface} and \eqref{eq:lowerSurface} and 
the scheme in Eq. \eqref{eq:zbimpl} for Eq. \eqref{eq:zbeq2} and $p_w$ as in \cite{Durand09b}. The difference between the approximations of $z_b$ in Eq. \eqref{eq:zint} and \eqref{eq:zbimpl} is of $\calO(\Delta t^2)$.

\section{Subgrid modeling around grounding line}
\label{sec:subgrid}

The basic idea of the subgrid method for the FS equations in this paper follows the GL parameterization for SSA in \cite{Seroussi14} and the analysis for FS in \cite{Schoof11}.
The GL is located at the position where the ice is on the ground and the flotation criterion is perfectly satisfied such that $\sigma_{\bn\bn}=-p_w$.
In the Stokes equations, the hydrostatic assumption may not be valid, so the exact GL position can not be determined by simply checking the total thickness of the ice $H$ against the depth below sea level $H_{bw}=-z_b$.
Instead, the flotation criterion is computed by comparing the water pressure with the normal stress component orthogonal to the boundary as indicated by the first order analysis in Sect. \ref{sec:GLsol}. The indicator is here defined by $\chi(x)=\sigma_{\bn\bn}+p_w$ which vanishes on the floating ice and is approximately $\tau_{22}-p+p_w$ and negative on the ground since the slope of the bedrock is small.

\begin{figure}[htbp]
\center
  \includegraphics[width=0.45\textwidth,page=1]{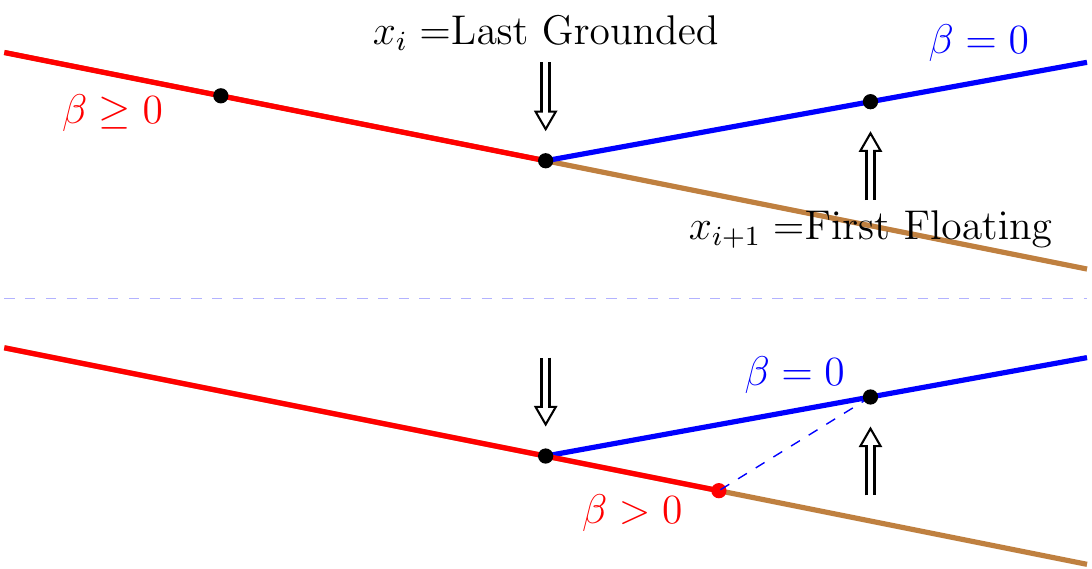} 
  \caption{Schematic figure of Grounding Line in case {\romannumeral 1}.
  Upper panel: the last grounded and first floating nodes as defined in Elmer/ICE. 
  Lower panel: linear interpolation to compute a more accurate position of the Grounding Line.}
  \label{fig:GL}
\end{figure}

\begin{figure}[htbp]
\center
  \includegraphics[width=0.45\textwidth,page=2]{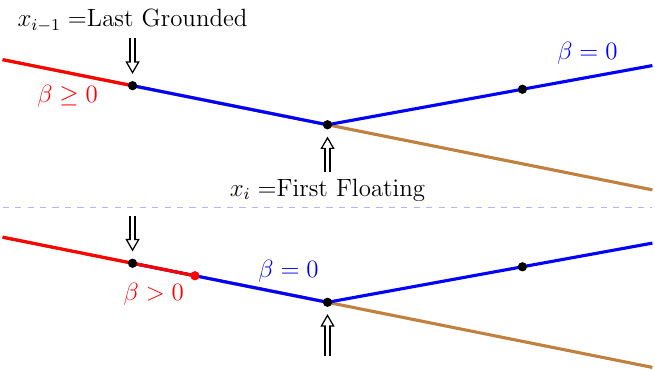} 
  \caption{Schematic figure of Grounding Line in case {\romannumeral 2}.
  Upper panel: the last grounded and first floating nodes as defined in Elmer/ICE. 
  Lower panel: linear interpolation to compute a more accurate position of the Grounding Line.}
  \label{fig:GL2}
\end{figure}

Typically, at the lower surface of the floating ice where $z_b(x,t)> b(x)$, as the blue line in Fig. \ref{fig:GL}, the boundary conditions are given by Eq. \eqref{eq:BCFI}, and where the ice is in contact with the bedrock, as the red line in Fig. \ref{fig:GL}, the boundary conditions are given by Eq. \eqref{eq:BCGI}. 
However, there is another case as shown in Fig. \ref{fig:GL2} when the net force at $x_i$ is pointing inward, namely $\sigma_{\bn\bn}(x_i)+p_w(x_i)>0$.
Then, the floating boundary condition Eq. \eqref{eq:BCFI} should be imposed up until the node $x_{i-1}$.
This can happen at some point due to the low spatial and temporal resolutions, but the node $x_i$ will move upward as long as $\bu\cdot\bn<0$, or the net force switches signs and the condition transforms into the case in Fig. \ref{fig:GL} when $\sigma_{\bn\bn}(x_i)+p_w(x_i)<0$.
Denote the situation in Fig. \ref{fig:GL} by case {\romannumeral 1}, and the one in Fig. \ref{fig:GL2} by case {\romannumeral 2}.
We call the node `grounded' when it is in contact with the bedrock with net force from the ice pointing outward ($\sigma_{\bn\bn}+p_w<0$), and `floating' when the net force is pointing inward ($\sigma_{\bn\bn}+p_w\geq0$).
The element which contains both grounded and floating nodes is called the GL element and the grounded node in it is called the last grounded node and the floating one is called the first floating node.

In coarse meshes, the true position of the GL is generally not in one of the nodes, but usually between the last grounded and the first floating nodes. 
Instead of refining the mesh around GL, which would lead to very small time steps for stability reasons, we will here introduce a subgrid model for the GL element.

We let $\chi(x)=\sigma_{\bn\bn}(x)+p_w(x)$ and assume that it is linear as in Eq. \eqref{eq:chidef} to determine the position of the GL, $x_{GL}$, in the GL element. 
In case {\romannumeral 2}, the GL is located between $x_{i-1}$ and $x_i$ even though the whole element $[x_{i-1},x_i]$ is geometrically grounded.
The equation $\chi(x_{GL})=0$ is solved by linear interpolation between $\chi(x_{i-1})<0$ and $\chi(x_i)>0$ yielding a unique solution satisfying $x_{i-1}<x_{GL}<x_i$, depicted as the red dot in the lower panel of Fig. \ref{fig:GL2}.

There is a more complicated situation in case {\romannumeral 1}, where $\chi(x_i)<0$ but $\chi(x_{i+1})=0$ due to the floating boundary condition.
A correction of $\chi$ is made by using $\tilde{\chi}(x)=\sigma_{\bn\bn}(x)+p_b(x)$ where $p_b(x)=-\rho_w gb(x)$ is the water pressure on the bedrock.
For $x>x_i$, we have $b(x)<z_b(x)$ and $p_b(x)>p_w(x)$.
Therefore, $\tilde{\chi}(x_{i+1})>\chi(x_{i+1})=0$ and  $\tilde{\chi}(x_i)=\chi(x_i)<0$.
Then, a linear interpolation between $\tilde{\chi}(x_i)$ and $\tilde{\chi}(x_{i+1})$ guarantees a unique solution of $\tilde{\chi}(x_{GL})=0$ in the GL element $[x_i,x_{i+1}]$, see Fig. \ref{fig:GL}.
In case {\romannumeral 2}, $p_b$ can also be used since $p_b(x)=p_w(x)$ as long as the element is on the bedrock.

Conceptually, the linear interpolation of the function $\tilde{\chi}(x)$ can be considered separately by looking at the two linear functions $\sigma_{\bn\bn}(x)$ and $p_b(x)$. 
As the GL always rests on the bedrock, $p_b(x_{GL})=p_w(x_{GL})$ is actually an exact representation of the water pressure imposed on the ice at GL, although geometrically $z_b(x_{GL})$ may not coincide with $b(x_{GL})$, especially on coarse meshes.
This also leads to the fact that the interpolated normal stress $\sigma_{\bn\bn}(x_{GL},z_b(x_{GL}))$ is a first order approximation of the normal stress at the exact GL position $(x_{GL},b(x_{GL}))$.

This correction is not necessary when the GL is advancing since the implicit treatment of the bottom surface is equivalent to additional water pressure at the stress boundary as discussed in Sect. \ref{sec:updlower}.

After the GL position is determined, the domains $\Gamma_{bg}$ and $\Gamma_{bf}$ are separated at $x_{GL}$ as in Eq. \eqref{eq:Nitscheint} and the integrals are calculated with a high-order integration scheme as in \cite{Seroussi14} to achieve a better resolution within the element shown in Figures \ref{fig:GL} and \ref{fig:GL2}.
For a smoother transition of $\beta$ at $GL$, the slip coefficient is multiplied by 1/2 at the whole GL element before integrating using the high order scheme.


The penalty term from Nitsche's method restricts the motion of the element in the normal direction. It should only be imposed on the element which is fully on the ground.
On the contrary, in case {\romannumeral 1}, the GL element $[x_i,x_{i+1}]$ is not in contact with the bedrock as in Fig. \ref{fig:GL}, so only the floating boundary condition should be used on the element $[x_i,x_{i+1}]$.
Additionally, the implicit representation of the bottom surface in Eq. \eqref{eq:zbimpl} also implies that the case {\romannumeral 2}\, with retreating GL should be merged to case {\romannumeral 1}\, since the surface is leaving the bedrock and the normal velocity should not be forced to zero.
To summarize, Nitsche's penalty term should be imposed on all the fully grounded elements and partially on the GL element in the advance phase.

Equations (\ref{eq:FS}), (\ref{eq:freeSurface}), and (\ref{eq:lowerSurface}) form a system of coupled nonlinear equations. They are solved in the same manner as in Elmer/ICE v.8.3.
The $x_{GL}$ position is determined dynamically within every nonlinear iteration when solving the FS equations and the high order integrations are based on the current $x_{GL}$.
The nonlinear FS is solved with fixed-point iterations to $10^{-5}$ relative error with a limit of maximal 25 nonlinear iterations and the grounded condition is set if the distance between of the bottom surface and the bedrock is smaller than $10^{-3}$~m.

\section{Results} \label{sec:results}

The numerical experiments follow the MISMIP benchmark \cite{MISMIP} and comparison is made with the results in \cite{gagliardini2016impact}.
Using the experiment MISMIP 3a, the setups are exactly the same as in the advancing and retreating simulations in \cite{gagliardini2016impact}.
The experiments are run with spatial resolutions of $\Delta x=4$~km, 2~km and 1~km with 20 vertical extruded layers.
The timestep is $\Delta t=0.125$~year for all the three resolutions to eliminate time discretization errors when comparing different spatial resolutions.

\begin{figure}[htbp]
\center
  \includegraphics[width=0.4\textwidth]{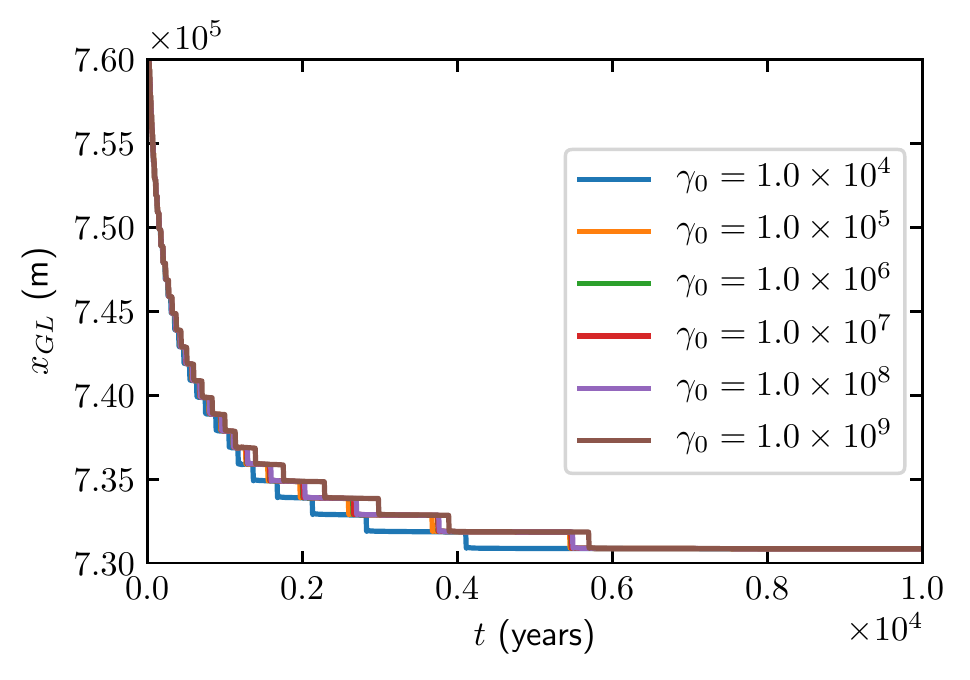} 
  \caption{The MISMIP 3a retreat experiment with $\Delta x=1000$~m for different choices of $\gamma_0$ in
    the time interval $[0,10000]$ years.}
\label{fig:gammas}
\end{figure}

The dependence on $\gamma_0$ for the retreating ice is shown in Fig. \ref{fig:gammas} with $\gamma_0$ between $10^4$ and $10^9$.
The estimated GL positions do not vary with different choices of $\gamma_0$ from $10^5$ to $10^8$ which suggests a suitable range of $\gamma_0$.
If $\gamma_0$ is too small ($\gamma_0\ll10^4$), oscillations appear in the estimated GL positions. 
If $\gamma_0$ is too large ($\gamma_0\gg10^8$), then more nonlinear iterations are needed for each time step.
The same dependency of $\gamma_0$ is observed for the advance experiments and for different mesh resolutions as well.
For the remaining experiments, we fix $\gamma_0=10^6$.

\begin{figure}[htbp]
\center
  \includegraphics[width=0.45\textwidth]{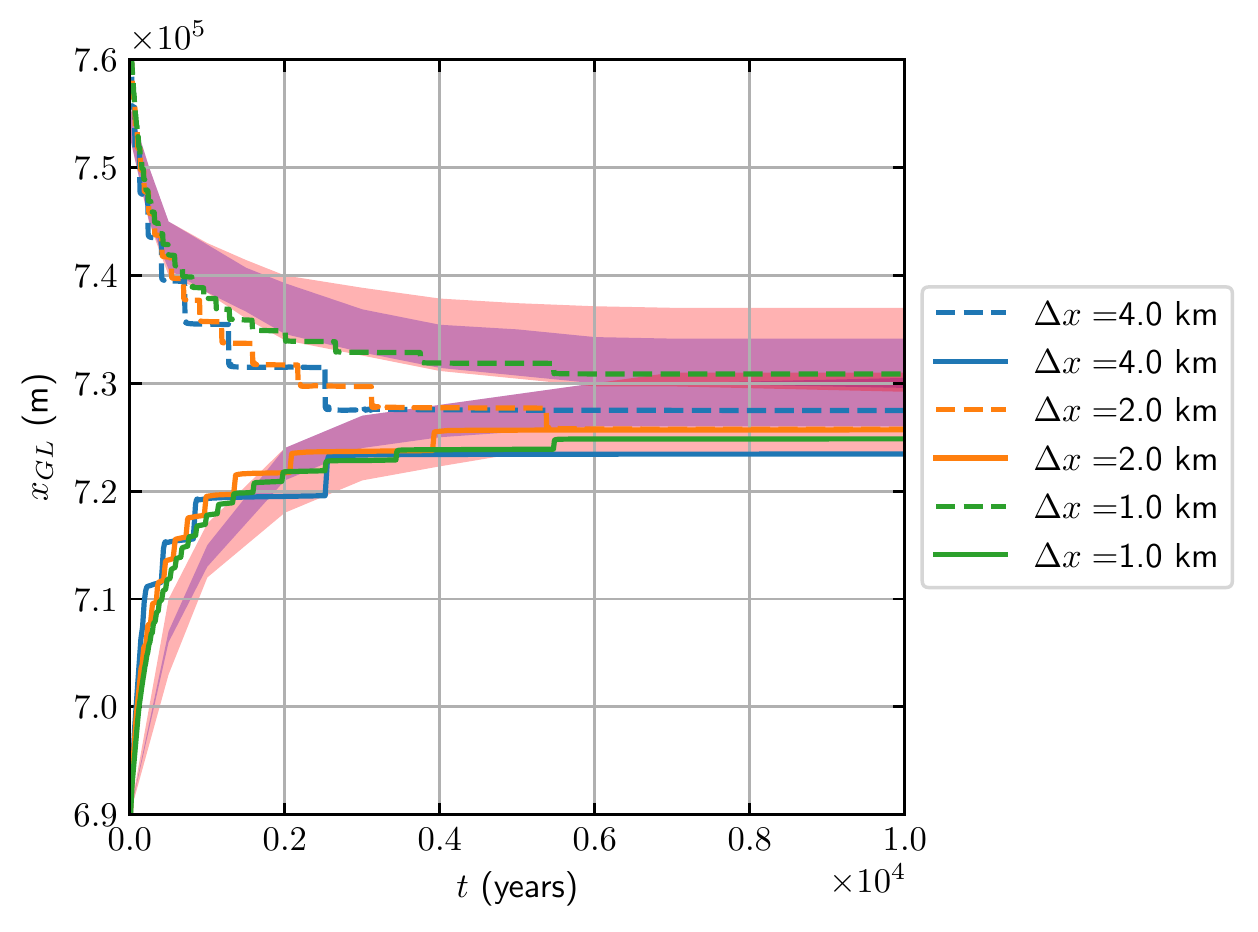} 
  \caption{The MISMIP 3a experiments for the GL position when $t\in[0, 10000]$ with $\Delta x=4000, 2000$ and $1000$~m for the advance (solid) and retreat (dashed) phases. 
  The shaded regions indicate the range of the results in \cite{gagliardini2016impact} with $\Delta x=50$~m in red and $\Delta x=25$~m in blue.   }
\label{fig:MISMIP3}
\end{figure}

The GL position during 10000 years in the advance and retreat phases are displayed in Fig. \ref{fig:MISMIP3} for different mesh sizes.
The range of the results from \cite{gagliardini2016impact} with mesh resolutions $\Delta x=25$ and $50$~m are shown as background shaded regions with colors purple and pink.
We achieve similar GL migration results both for the advance and retreat experiments with at least 20 times larger mesh sizes.

\begin{figure}[htbp]
\center
  \includegraphics[width=0.45\textwidth]{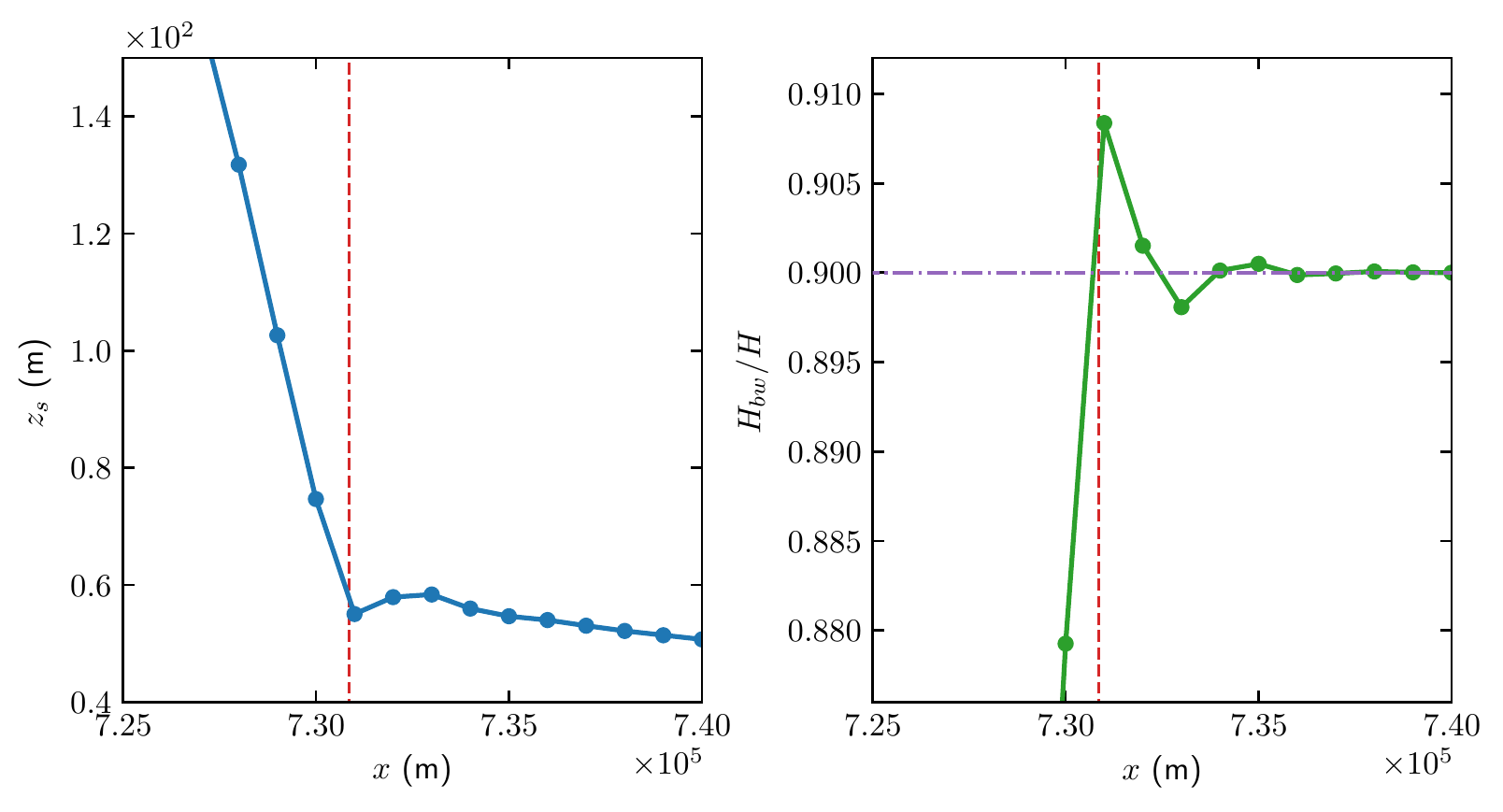} 
  \caption{Details of the solutions for the retreat experiment with $\Delta x=1$~km after 10000 years. 
  The solid dots represent the nodes of the elements and the vertical, red, dashed lines indicate the GL position.
  \emph{Left panel}: The oscillations at top surface near GL.
  \emph{Right panel}: The flotation criterion is evaluated by $H_{bw}/H$. The ratio between $\rho/\rho_w$ is drawn in a horizontal, purple, dash-dotted line. 
  }
\label{fig:surfOsc}
\end{figure}

We observed oscillations at the top surface near the GL in all the experiments as expected from \cite{Durand09b, Schoof11}.
A zoom-in plot of the surface elevation with $\Delta x=1$~km at $t=10000$ years is shown to the left in Fig. \ref{fig:surfOsc}, where the red dashed line indicates the estimated GL position. 
Obviously, the estimated GL position does not coincide with any nodes even at the steady state.

The ratio between the thickness below sea level $H_{bw}$ and the ice thickness $H$ is shown in Fig. \ref{fig:surfOsc}.
The horizontal, purple, dash-dotted line indicates the ratio of $\rho/\rho_w$ and the estimated GL is located at the red, dashed line.
This result confirms that the hydrostatic assumption $H\rho=H_{bw}\rho_w$ is not valid in the FS equations for $x>x_{GL}$ close to the GL and at the GL position, cf. \cite{Durand09b, Schoof11}. For $x<x_{GL}$ we have that $H_{bw}/H<\rho/\rho_w$ since $H_{bw}$ decreases and $H$ increases. The conclusion from numerical experiments in \cite{van2018dynamically} is that the hydrostatic assumption and the SSA equations approximate the FS equations well for the floating ice beginning at a short distance away from the GL.

\begin{figure}[htbp]
\center
  \includegraphics[width=0.45\textwidth]{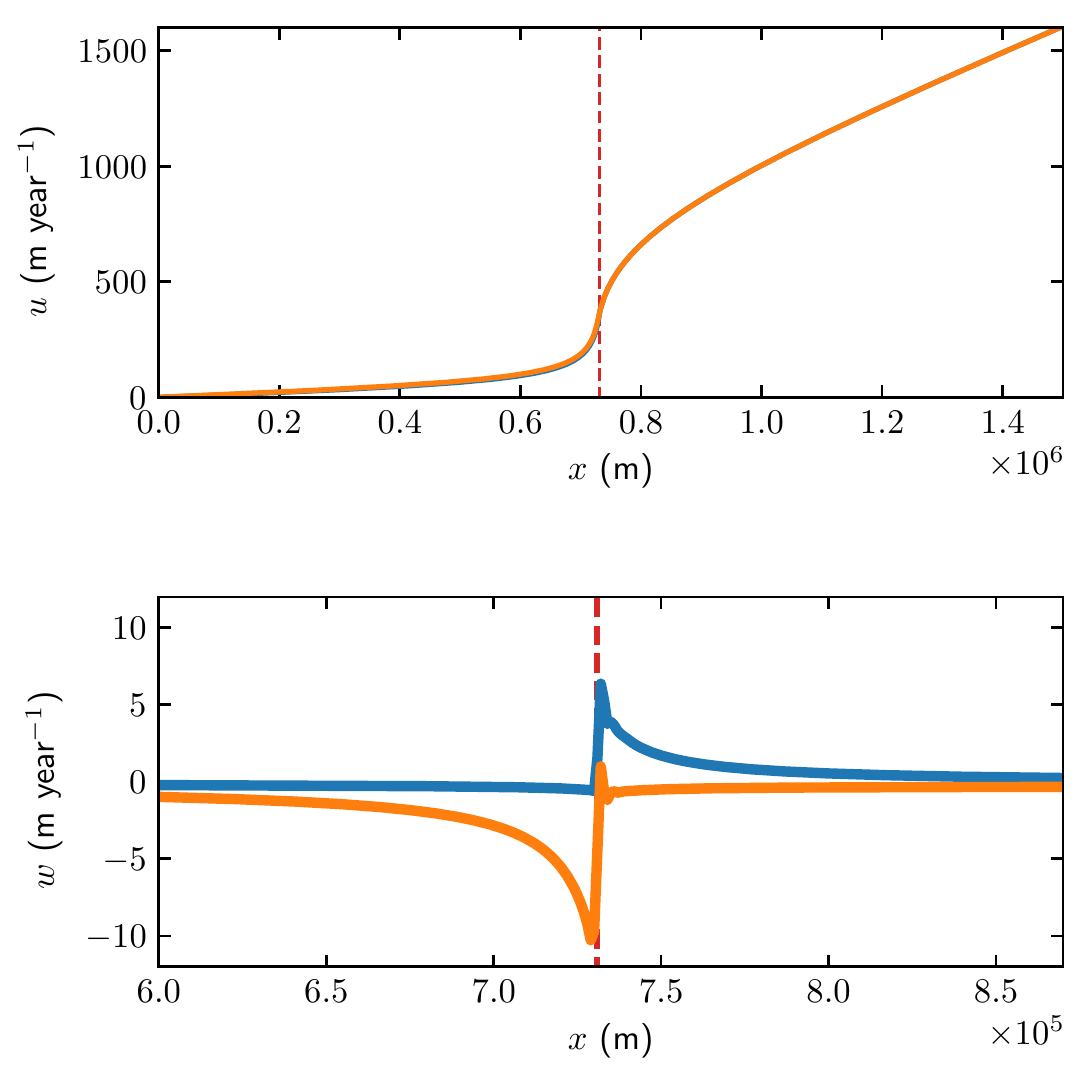} 
  \caption{The velocities $u$ (upper panel) and $w$ (lower panel) on the top (orange) and bottom (blue) surface of the ice in the retreat experiment with $\Delta x=1$~km after 10000 years. The red, dashed line indicates the GL position. The vertical velocity $w$ is zoomed-in close to the GL.}
\label{fig:velocity}
\end{figure}

The top and bottom surface velocity solutions from the retreat experiment are shown in Fig. \ref{fig:velocity} with $\Delta x=1$~km after 10000 years.
The horizontal velocities on the two surfaces are similar with negligibly small differences on the floating ice.
The vertical velocities $w$ on the top (orange line) and bottom surface (blue line) at the GL are almost discontinuous as analyzed in \cite{Schoof11}.
With the subgrid method, the rapid variation is resolved by the 1 km mesh size.

\section{Discussion}\label{sec:disc}

Seroussi et al \cite{Seroussi14} describe four different subgrid models for the friction in SSA and evaluate them in a FEM discretization on a triangulated, planar domain. 
The flotation criterion is applied
at the nodes of the triangles. Depending on how many of the nodes that are floating, the amount of friction in the triangle is determined. Also, a higher order polynomial integration 
over the triangles in FEM allows an inner structure in the triangular element.

Our method can be extended to a triangular mesh covering $\Gamma_b$ in the following way. 
The condition on $\chi$ in Eq. \eqref{eq:chidef} is applied on the edges of each triangle $\calT$ in the mesh. If $\chi<0$ in all three nodes 
then $\calT$ is grounded. If $\chi\ge 0$ in all nodes then $\calT$ is floating. The GL passes inside 
$\calT$ if $\chi$ has a different sign in one of the nodes. Then the GL crosses the two edges where
$\chi<0$ in one node and $\chi\ge 0$ in the other node. In this way, a continuous reconstruction of a piecewise linear GL is possible on $\Gamma_b$. 
The FEM approximation is modified in the same manner as in Sect.~\ref{sec:subgrid} with Nitsche's method. 

An alternative to subgrid modeling is to introduce dynamic adaptation of the mesh on $\Gamma_b$ with a refinement at the GL as in e.g. \cite{Cornford13, Drouet13, Gladstone10a}.
In general, a fine mesh is needed along the GL and in an area surrounding it. Since the GL moves long distances at least in simulations of palaeo-ice sheets, 
the adaptation should be dynamic, permit refinement and coarsening 
of the mesh, and be based on some estimate of the model inaccuracy. Furthermore, shorter timesteps are necessary for stability when the mesh size is smaller in a mesh adaptive  method. Introducing a time dependent mesh adaptivity into an existing code requires a substantial coding effort and will
increase the computational work considerably. Subgrid modeling is easier to implement and the increase in computing time is small.

\conclusions\label{sec:concl}

Subgrid models at the GL have been developed and tested in the SSA model for 2D flow in \cite{Gladstone10b} and for 3D flow in \cite{Seroussi14}, for the friction in the 
vertically integrated model BISICLES \cite{Cornford13} for 3D flow in \cite{Cornford16},
and for the PISM model mixing SIA with SSA in 3D in \cite{Feldmann14}. Here we propose a subgrid model in 2D for the FS equations implemented in Elmer/ICE that can be extended to 3D.
The mesh is static and the moving GL position within one element is determined by linear interpolation with an auxiliary function $\tilde{\chi}$ based on the theory in \cite{Schoof11}. 
Only in that element, the FEM discretization is modified.
 
The method is applied to the simulation of an ice sheet in 2D with an advancing GL and one with a retreating GL. The data for the tests are the same as in one of the MISMIP examples \cite{MISMIP} and in \cite{gagliardini2016impact}. Comparable results to \cite{gagliardini2016impact} are obtained with subgrid modeling with more than 20 times larger mesh sizes. A larger mesh size also allows a longer timestep for the time integration. Without further knowledge of the basal conditions and detailed models at the GL, solving $\tilde{\chi}(x)=0$ provides a good approximation of the GL position.

\codeavailability{  
The FS sub-grid model is implemented based on Elmer/ICE Version: 8. 3(Rev: f6bfdc9) with the scripts at \url{https://github.com/enigne/elmerfem/tree/NitscheMethodAdvance} and \url{https://github.com/enigne/elmerfem/tree/NitscheMethodRetreat}.
}

\authorcontribution{GC developed the model code and performed the simulations. GC and PL contributed to the theory of the paper. GC, PL and LvS contributed to the development of the method and the writing of the paper} 

\competinginterests{The authors declare that they have no conflict of interest.} 

\begin{acknowledgements}
This work has been supported by Nina Kirchner's Formas grant 2017-00665 and the Swedish e-Science initiative eSSENCE.
We are grateful to Thomas Zwinger for advise and help in the implementation of the subgrid model in Elmer/ICE.
The computations were performed on resources provided by the Swedish National Infrastructure for Computing (SNIC) at the PDC Center for High Performance Computing, KTH Royal Institute of Technology.
\end{acknowledgements}

\bibliographystyle{copernicus}
\bibliography{iceGL}

\end{document}